\documentclass[twocolumn,superscriptaddress,nobibnotes,aps,prd,showpacs,nofootinbib]{revtex4}
\usepackage{amsmath}
\usepackage{amsfonts}
\usepackage{amssymb}
\usepackage{graphicx}

\def\be{\begin{equation}}
\def\ee{\end{equation}}
\def\bea{\begin{eqnarray}}
\def\eea{\end{eqnarray}}

\begin{document}

\title{Gravitational induced particle production through a nonminimal
curvature-matter coupling}
\author{Tiberiu Harko}
\email{t.harko@ucl.ac.uk}
\affiliation{Department of Mathematics, University College London, Gower Street, London
WC1E 6BT, United Kingdom}
\author{Francisco S. N. Lobo}
\email{fslobo@fc.ul.pt}
\affiliation{Instituto de Astrof\'{\i}sica e Ci\^{e}ncias do Espa\c{c}o, Faculdade de
Ci\^encias da Universidade de Lisboa, Edif\'{\i}cio C8, Campo Grande,
P-1749-016 Lisbon, Portugal}
\author{Jos\'e P. Mimoso}
\email{jpmimoso@fc.ul.pt}
\affiliation{Instituto de Astrof\'{\i}sica e Ci\^{e}ncias do Espa\c{c}o, Faculdade de
Ci\^encias da Universidade de Lisboa, Edif\'{\i}cio C8, Campo Grande,
P-1749-016 Lisbon, Portugal}
\author{Diego Pav\'{o}n}
\email{diego.pavon@uab.es}
\affiliation{Departamento de F\'{\i}sica, Universidad Aut\'{o}noma de Barcelona, 08193
Bellaterra (Barcelona), Spain}

\begin{abstract}
We consider the possibility of a gravitationally induced particle production
through the mechanism of a nonminimal curvature-matter coupling. An interesting feature of this gravitational theory is that the divergence of the energy-momentum tensor is
nonzero. As a first step in our study we reformulate the model in terms of
an equivalent scalar-tensor theory, with two arbitrary potentials. By using
the formalism of open thermodynamic systems, we interpret the energy balance
equations in this gravitational theory from a thermodynamic point of view,
as describing irreversible matter creation processes. The particle number
creation rates, the creation pressure, and the entropy production rates are
explicitly obtained as functions of the scalar field and its potentials, as
well as of the matter Lagrangian. The temperature evolution laws of the
newly created particles are also obtained. The cosmological implications of
the model are briefly investigated, and it is shown that the late-time
cosmic acceleration may be due to particle creation processes. Furthermore,
it is also shown that due to the curvature--matter coupling, during the
cosmological evolution a large amount of comoving entropy is also produced.
\end{abstract}

\date{\today }
\maketitle


\section{Introduction}

The simplest explanation of the late-time accelerated expansion of the
Universe \cite{1,2} is to invoke a cosmological constant, $\Lambda$, which
can be associated to the vacuum energy \cite{Carroll:2000fy}. Despite the
excellent fit to observational data, the presence of $\Lambda$ suffers from
two serious drawbacks, namely, the cosmological constant problem and the
coincidence problem. However, it has recently been argued that the vacuum
energy is not constant but decays into other particle constituents \cite%
{Graef:2013iia}. Indeed, phenomenological models, with a variable
cosmological constant \cite{varyLambda}, have been proposed to address the
above problems. For instance, a simple and thermodynamically consistent
cosmology with a phenomenological model of quantum creation of radiation due
to vacuum decay was presented in \cite{Gunzig:1997tk}, where the
thermodynamics and Einstein's equations lead to an equation in which $H$ is
determined by the particle number $N$. It was shown that the evolution
equation for $H$ has a remarkably simple exact solution, in which a
non-adiabatic inflationary era exits smoothly to the radiation era, without
a reheating transition. In \cite{Lima:2008qy}, a new accelerating flat model
without dark energy that is fully dominated by cold dark matter (CDM) was
investigated. It was shown that the number of CDM particles is not conserved
and the present accelerating stage is a consequence of the negative pressure
describing the irreversible process of gravitational particle creation.

In \cite{Graef:2013iia}, the correspondence between cosmological models
powered by a decaying vacuum energy density and gravitationally induced
particle production was explored. Although being physically different, it
was shown that under certain conditions both classes of cosmologies can
exhibit the same dynamical and thermodynamical behavior. By using current
type Ia supernovae data, recent estimates of the cosmic microwave background
shift parameter and baryon acoustic oscillations measurements, the authors
performed a statistical analysis to test the observational viability of the
models and the best-fit of the free parameters was also obtained.
Furthermore, the particle production cosmologies (and the associated
decaying $\Lambda(t)$-models) were modelled in the framework of field theory
by a phenomenological scalar field model.

In this context, a new cosmic scenario with gravitationally induced particle
creation was proposed \cite{Lima:2014qpa}, where the Universe evolves from
an early to a late time de Sitter era, with the recent accelerating phase
driven only by the negative creation pressure associated with the cold dark
matter component. The model can be interpreted as an attempt to reduce the
so-called cosmic sector (dark matter plus dark energy) and relate the two
cosmic accelerating phases (early and late time de Sitter expansions). A
detailed thermodynamic analysis including possible quantum corrections was
also carried out. For a very wide range of the free parameters, it was found
that the model presents the expected behavior of an ordinary macroscopic
system in the sense that it approaches thermodynamic equilibrium in the long
run (i.e., as it nears the second de Sitter phase). Moreover, an upper bound
was found for the Gibbons-Hawking temperature of the primordial de Sitter
phase \cite{Mim}. Finally, when confronted with the recent observational data, the
current `quasi'-de Sitter era, as predicted by the model, it was verified to
pass the cosmic background tests very comfortably.

In this work, we consider an alternative mechanism for the gravitational
particle production, namely, through a nonminimal curvature-matter coupling,
in modified theories of gravity. A general property of these theories \cite
{Bertolami:2007gv,fRL1,fRL2, fRL3,fRL4,fT}, is the non-conservation of the
energy-momentum tensor (for a recent review of modified gravity models with
curvature-matter coupling see \cite{gal}). Thus, the coupling between the
matter and the higher derivative curvature terms may be interpreted as an
exchange of energy and momentum between both. This latter mechanism induces
a gravitational particle production. We note that the generalized energy
balance equations in these gravitational theories have been interpreted from
a thermodynamic point of view as describing irreversible matter creation
processes in \cite{therm1}. Thus, the coupling between matter and geometry
generates an irreversible energy flow from the gravitational field to newly
created matter constituents, with the second law of thermodynamics requiring
that the geometric curvature transforms into matter.

Here we extend and refine the analysis
initiated in \cite{therm1} by investigating in detail the thermodynamic
interpretation of the curvature-matter gravitational coupling for the
so-called linear version of the $f(R,L_m)$ gravity theory, where $R$ is the Ricci scalar and $L_m$ is the matter Lagrangian, which we denote
as $Lf(R,L_m)$ theory, with the gravitational Lagrangian given by $f_1(R)/2+%
\left[1+\lambda f_2(R)\right]{\ L}_{m}$, where $f_1(R)$ and $f_2(R)$ are
arbitrary functions of $R$, and $\lambda $ is a coupling constant. We note that in the $Lf(R,L_m)$ theory, the gravitational action is {\it linear in the matter Lagrangian $L_m$}, and not in the Ricci scalar $R$.  As a first step in our
study we introduce the equivalent scalar-tensor description of the theory
\cite{Faraoni:2007sn}, in which the action is equivalent to a Brans-Dicke type
theory, with a single scalar field $\psi $, a vanishing Brans-Dicke
parameter $\omega $, and a coupling $U(\psi )$ between the scalar field and
matter.

By using the formalism of open thermodynamic systems \cite{Pri0,Pri,Cal,Lima},
we interpret the energy balance equation of the theory as describing a
matter creation process. Indeed, the irreversible thermodynamics of open
systems, and its implications for cosmology have been extensively analyzed
\cite{ref}. Here, we obtain the equivalent particle number creation rates,
the creation pressure and the entropy production rates as functions of the
scalar field, of the two scalar potentials, and of the matter Lagrangian,
respectively. The temperature evolution of the newly created particles is
also obtained. Due to the curvature--matter coupling, during the
cosmological evolution a large amount of comoving entropy could be produced.
The cosmological implications of the theory in its scalar-tensor
representation are also investigated.

The present paper is organized as follows. In Section \ref{sect2}, we present
the action and the field equations of the modified gravity model with a
linear curvature-matter coupling in both their standard and scalar-tensor
representations. The thermodynamic interpretation of the theory is developed in
Section~\ref{sect3}, where the particle creation rates, the creation
pressure and the entropy production are analyzed in detail. In Section~\ref{sect4}, the
cosmological implications of the theory in its scalar-tensor representation are
considered. In Section~\ref{sectn}, the behavior of the  entropy of the Universe in the
$Lf\left(R,L_m\right)$ gravity theory, with the horizon entropy included, is
analyzed. We discuss and conclude our results in
Section~\ref{sect5}.

\section{Nonminimal curvature-matter coupling}
\label{sect2}

\subsection{General formalism}

The action of $f(R)$ gravity can be generalized with the introduction of a
linear nonminimal coupling between matter and curvature. The corresponding
gravitational theory, a particular case of the general $f(R,L_m)$ theory
\cite{fRL2}, and which we denote as $Lf\left(R,L_m\right)$, has the action given by \cite{Bertolami:2007gv},
\begin{equation}
S=\int \left\{\frac{1}{2}f_1(R)+\left[1+\lambda f_2(R)\right]{\ L}%
_{m}\left(g_{\mu \nu},\Phi\right)\right\} \sqrt{-g}\;d^{4}x~
\label{actionlinear}
\end{equation}
where the factors $f_i(R)$ (with $i=1,2$) are arbitrary functions of the
Ricci scalar $R$. The coupling constant $\lambda$ determines the strength of
the interaction between $f_2(R)$ and the matter Lagrangian. ${L}_{m}$ is the
matter Lagrangian density, which is a function of the metric $g_{\mu \nu}$
and of the matter fields $\Phi $.

Now, varying the action with respect to the metric $g_{\mu \nu }$ provides
the following field equations:
\begin{equation}
\Theta R_{\mu \nu }-\frac{1}{2}f_{1}(R)g_{\mu \nu }+\hat{P}_{\mu \nu }\Theta
=[1+\lambda f_{2}(R)]T_{\mu \nu },  \label{field1a}
\end{equation}%
where $F_{i}(R)=f_{i}^{\prime }(R)$, $i=1,2$, and the prime represents a
derivative with respect to the scalar curvature $R$. We have defined
\begin{equation}
\Theta =F_{1}(R)+2\lambda F_{2}(R)L_{m},
\end{equation}%
and
\begin{equation}
\hat{P}_{\mu \nu }=(g_{\mu \nu }\square -\nabla _{\mu }\nabla _{\nu }),
\label{id1}
\end{equation}%
for notational simplicity.
The matter energy-momentum tensor is defined as
\begin{equation}
T_{\mu \nu }=-\frac{2}{\sqrt{-g}}\frac{\delta (\sqrt{-g}\,{L}_{m})}{\delta
(g^{\mu \nu })} \,.  \label{EMTdef2}
\end{equation}

An important property of any gravitational theory is its stability with respect to local perturbations. In the standard $f(R)$ gravity, a fatal instability (the Dolgov-Kawasaki instability)  appears once  the condition $f''(R) < 0$ \cite{ins}. The instability develops on time scales of the order of  $10^{−26}$ s. The stability properties of the gravitational models described by the action given by Eq.~(\ref{actionlinear}) were studied in \cite{Farstab}, and it turns out that the corresponding stability criterion is $f^{''}_1(R)+2\lambda f^{''}_2(R)>0$.  In order to obtain the stability condition one expands
the parameters of the model as the sum of a background field with constant curvature, and a small perturbation, so that $R = R_0 + R_1$, $T = T_0 + T_1$, $f_1(R)=R_0+R_1+\epsilon \phi \left(R_0\right)+\epsilon \phi '\left(R_0\right)R_1+...$, $f_1'(R)=1+\epsilon \phi '\left(R_0\right)+\epsilon \phi ''\left(R_0\right)R_1+...$. With the use of the linearized field equations we obtain the stability condition $\epsilon \phi ''(R)+2\lambda f_2^{''}(R)>0$ \cite{Farstab}, generalizing the stability condition $f''(R)=\epsilon \phi ''(R)>0$, found in $f(R)$ gravity \cite{ins}. From a physical point of view the reason for stability is once the stability conditions are satisfied  the effective mass $m_{eff}$ of the dynamical degree
of freedom associated to the small perturbation $R_1$ of the background curvature is non-negative.

A general property of these nonminimal curvature-matter coupling theories is
the non-conservation of the energy-momentum tensor. This can be easily
verified by taking into account the covariant derivative of the field Eq.~(%
\ref{field1a}), the Bianchi identities, $\nabla^\mu G_{\mu\nu}=0$, and the
following identity,
\begin{equation}  \label{id}
\mbox{$(\square\nabla_\nu
-\nabla_\nu\square)F_i=R_{\mu\nu}\,\nabla^\mu F_i$},
\end{equation}
which then implies the following relationship:
\begin{equation}
\nabla^\mu T_{\mu \nu}=\frac{\lambda F_2}{1+\lambda f_2}\left[%
g_{\mu\nu}{L}_m- T_{\mu \nu }\right]\nabla^\mu R .  \label{cons1}
\end{equation}
Note that in the absence of the coupling, $\lambda=0$, one obtains
the conservation of the energy-momentum tensor \cite{Koivisto:2005yk}, which
can also be verified through the diffeomorphism invariance of the matter
part of the action. The conservation of the energy-momentum tensor also
follows from Eq.~(\ref{cons1}), if $f_2(R)$ is a constant or the matter
Lagrangian is not an explicit function of the metric.

In order to test the motion in our model, we consider for the
energy-momentum tensor of matter a perfect fluid
\begin{equation}  \label{emt}
T_{\mu \nu }=\left( \rho +p\right) U_{\mu }U_{\nu }-pg_{\mu \nu },
\end{equation}
where $\rho$ is the total energy density and $p$, the pressure,
respectively. The four-velocity, $U_{\mu }$, satisfies the conditions $%
U_{\mu }U^{\mu }=1$ and $\nabla _{\nu }U^{\mu }U_{\mu }=0$. We also
introduce the projection operator $h_{\mu \lambda }=g_{\mu \lambda }-U_{\mu
}U_{\lambda }$ from which one obtains $h_{\mu \lambda }U^{\mu }=0$.

From Eq.~(\ref{cons1}), we deduce the equation of motion for a fluid
element:
\begin{equation}
\frac{D U^{\alpha }}{ds} \equiv \frac{dU^{\alpha }}{ds}+\Gamma _{\mu \nu
}^{\alpha }U^{\mu }U^{\nu }=f^{\alpha }~  \label{eq1}
\end{equation}
where the extra force is given by:
\begin{eqnarray}  \label{force}
f^{\alpha }&=&\frac{1}{\rho +p}\left[\frac{\lambda F_2}{1+\lambda f_2}\left({%
L}_m + p\right)\nabla_\nu R+\nabla_\nu p \right] h^{\alpha \nu }.
\end{eqnarray}

An intriguing feature is that the extra force depends on the form of the
Lagrangian density. Note that considering the Lagrangian density $L_m = - p$, where $p$ is the pressure, the contribution of the nonminimal curvature-matter vanishes \cite{Bertolami:2008ab}.
It has been argued that this is not the unique choice for the matter
Lagrangian density and that more natural forms for $L_m$, such as $L_m =
\rho $, do not imply the vanishing of the extra-force. Indeed, in the
presence of the nonminimal coupling, they give rise to two distinct theories
with different predictions~\cite{Faraoni:2009rk, Bertolami:2013uwl}.

\subsection{Scalar-tensor representation of the linear curvature-matter coupling}

As has been shown in \cite{equiv1, equiv2, equiv3, equiv4}, $f(R)$ gravity
is equivalent to a scalar-tensor theory. In this context, the equivalence
between the modified gravity models to a linear curvature-matter coupling
and scalar-tensor gravity models was also established in \cite%
{Faraoni:2007sn}. More specifically, it was shown that the action given by
Eq.~(\ref{actionlinear}) is equivalent with a two-potential scalar-tensor
Brans-Dicke type theory, with a single scalar field, a vanishing Brans-Dicke
parameter $\omega$, and an unusual coupling of the second potential $U(\psi)$
of the theory to matter.

As a first step in the scalar-tensor formulation of the theory we introduce
a new field $\phi$, and reformulate the action~(\ref{actionlinear}) as
\begin{equation}  \label{100}
S=\int d^4x \sqrt{-g} \Bigg\{ \frac{f_1(\phi)}{2} +\frac{1}{2} \, \frac{df_1%
}{d\phi} \left( R-\phi \right) +\left[ 1+\lambda f_2( \phi) \right] L_m %
\Bigg\}.
\end{equation}
Next, we introduce the second field $\psi(\phi) \equiv f_1^{\prime}(\phi) $
(with a prime denoting a differentiation with respect to $\phi$), and thus
we obtain for the action the final expression
\begin{equation}  \label{300}
S=\int d^4x \sqrt{-g} \left[ \frac{\psi R }{2} -V(\psi)\, +U(\psi) L_m %
\right] ,
\end{equation}
where the two potentials $V(\psi)$ and $U(\psi)$ of the theory are defined
as
\begin{equation}  \label{400}
V(\psi) = \frac{\phi(\psi) f_1^{\prime }\left[ \phi (\psi ) \right] -f_1%
\left[ \phi( \psi ) \right] }{2},
\end{equation}
and
\begin{equation}  \label{500}
U( \psi) = 1+\lambda f_2\left[ \phi( \psi ) \right] ,
\end{equation}
respectively. The function $\phi (\psi)$ must be obtained by inverting $%
\psi(\phi) \equiv f_1^{\prime }(\phi) $. The actions~(\ref{actionlinear})
and (\ref{300}) are equivalent when $f_1^{\prime \prime }(R) \neq 0$ \cite%
{Faraoni:2007sn}, similarly to the case of pure $f(R)$ gravity \cite%
{equiv1,equiv2,equiv3,equiv4}.

The action given by Eq.~(\ref{300}) can be written, via a conformal rescaling $g_{\mu \nu}=\exp \left(-\alpha \phi /2\right)\hat{g}_{\mu \nu}$, $\alpha ={\rm constant}$,  as a four-dimensional dilaton gravity whose action, in the ``Einstein frame'', has the form \cite{Farstab}
\be
S_E=\int{d^4x\sqrt{-g}\left(\frac{\hat{R}}{2}-\nabla _{\mu}\phi \nabla ^{\mu} \phi -e^{-\alpha \phi }L_m\right)}.
\ee

In the  Einstein frame representation of the modified gravity with linear coupling between matter and geometry the extra force is due to the coupling between the matter Lagrangian and the Brans-Dicke-like scalar $\phi$. On the other hand it is important to mention that in the $Lf\left(R,L_m\right)$ theory there is no scaling of units with some powers of the conformal factor of the conformal transformation \cite{Farstab}.  For this reason it is impossible to reduce $Lf\left(R,L_m\right)$ gravity to a standard scalar-tensor theory, or to find a string gravity equivalent.

By varying the action (\ref{300}) with respect to $g_{\mu \nu}$ provides the
gravitational field equations of the scalar-tensor theory as
\begin{equation}  \label{14}
\hspace{-0.1cm}\psi \left(R_{\mu \nu}-\frac{1}{2}g_{\mu \nu}R\right)+\hat{P}%
_{\mu \nu}\psi =U(\psi)T_{\mu \nu}+V(\psi)g_{\mu \nu},
\end{equation}
while the variation of the action with respect to the field $\psi$ gives the
relation
\begin{equation}  \label{15}
\frac{R}{2}-V^{\prime }(\psi)+U^{\prime }(\psi)L_m=0.
\end{equation}

The contraction of the field equation Eq.~(\ref{14}) yields the scalar
relation
\begin{equation}
-\psi R+3\square \psi =U(\psi )T+4V(\psi ),  \label{16}
\end{equation}%
where $T=T_{\mu }^{\mu }$ is the trace of the energy-momentum tensor. By
combining Eqs.~(\ref{15}) and (\ref{16}) we obtain the field equation of the
field $\psi $ as
\begin{equation}  \label{square}
\square \psi =\frac{1}{3}U(\psi )T+\frac{4}{3}V(\psi )+\frac{2}{3}\psi
V^{\prime }(\psi )-\frac{2}{3}\psi U^{\prime }(\psi )L_{m}.
\end{equation}

By eliminating the term $\square \psi $, and taking into account Eq.~(\ref{square}) the field equations (\ref{14}) provides
\begin{eqnarray}  \label{eqff}
&&\psi \left(R_{\mu \nu}-\frac{1}{2}g_{\mu \nu}R\right)-\nabla _{\mu}\nabla
_{\nu }\psi=U(\psi)\Bigg(T_{\mu \nu}- \frac{1}{3}g_{\mu \nu}T\Bigg)  \notag
\\
&&-\left[\frac{1}{3}V(\psi)+\frac{2}{3}\psi V^{\prime }(\psi)-\frac{2}{3}\psi
U^{\prime }(\psi)L_m\right]g_{\mu \nu}.
\end{eqnarray}
Now, taking the covariant divergence of the field equation (\ref{14}),
with the use of Eq. (\ref{id}), we obtain first
\begin{eqnarray}
-\left[ \frac{R}{2}+V^{\prime }\left( \psi \right) \right] \nabla _{\nu
}\psi &=&U^{\prime }\left( \psi \right) \nabla _{\mu }\psi T_{\nu }^{\mu }
\notag \\
&& + U\left( \psi \right) \nabla _{\mu }T_{\nu }^{\mu }.
\end{eqnarray}
Then, by eliminating $R/2$ with the help of Eq. (\ref{15}) we obtain for the
divergence of the energy-momentum tensor
\begin{equation}
\nabla _{\mu }T_{\nu }^{\mu }=-\left[ \nabla _{\mu }\ln U\left( \psi \right) %
\right] T_{\nu }^{\mu }-\frac{2V^{\prime }\left( \psi \right) -U^{\prime
}(\psi )L_{m}}{U(\psi )}\nabla _{\nu }\psi .  \label{17}
\end{equation}

Equation~(\ref{17}) allows the formulation of the energy and momentum
balance equations in the scalar-tensor representation of the modified
theory of gravity with a linear coupling between matter and geometry. By
assuming that the energy-momentum tensor has the perfect fluid form given by
Eq.~(\ref{emt}), then Eq.~(\ref{17}) can be written in the equivalent form
\begin{eqnarray}
\left( \nabla ^{\mu }\rho +\nabla ^{\mu }p\right) U_{\mu }U_{\nu
}+\left(\rho +p\right) U_{\nu }\nabla ^{\mu }U_{\mu } -\nabla ^{\mu
}pg_{\mu\nu }  \notag \\
+\left( \rho +p\right) U_{\mu }\nabla ^{\mu }U_{\nu }+\nabla ^{\mu }\left[
\ln U\left( \psi \right) \right] T_{\mu \nu }  \notag \\
+\frac{2V^{\prime }\left( \psi \right) -U^{\prime }(\psi )L_{m}}{U(\psi )}\;
\nabla _{\nu }\psi =0.  \label{cem1}
\end{eqnarray}

By multiplying Eq.~(\ref{cem1}) with $U^{\nu }$ we obtain the energy balance
equation in the scalar-tensor representation of the linear curvature-matter
coupling given by
\begin{equation}  \label{eneq}
\dot{\rho}+3H(\rho +p)+\rho \frac{d}{ds}\ln U(\psi )+\frac{2V^{\prime
}\left( \psi \right) -U^{\prime }(\psi )L_{m}}{U(\psi )}\dot{\psi}=0,
\end{equation}%
where we have introduced the Hubble function $H=(1/3)\nabla ^{\mu }U_{\mu }$%
, and we have denoted $\dot{}=U^{\mu }\nabla _{\mu }=d/d{s}$, respectively,
where $d{s}$ is the line element corresponding to the metric $g_{\mu \nu }$,
$d{s}^{2}=g_{\mu \nu }dx^{\mu }dx^{\nu }$. After acting on Eq.~(\ref{cem1})
with the projection operator $h^{\nu \lambda }$, provides the momentum
balance equation for a perfect fluid as
\begin{eqnarray}  \label{force1}
U^{\mu }\nabla _{\mu }U^{\alpha }=\frac{d^{2}x^{\alpha }}{ds^{2}}+\Gamma
_{\mu \nu }^{\alpha }U^{\mu }U^{\nu }-f^\alpha =0
\end{eqnarray}
where the extra-force is given by
\begin{equation}
f^\alpha = h^{\mu \alpha }\Bigg[p\nabla _{\mu
}\ln U(\psi ) - \frac{2V^{\prime }\left( \psi \right) -U^{\prime }(\psi )L_{m}}{U(\psi )}%
\nabla _{\mu }\psi \Bigg] .
\end{equation}

It is important to note that in the scalar-tensor representation of modified
gravity with a linear curvature-matter coupling, the extra-force acting on
test fluids is non-zero independently of the choice for the matter
Lagrangian.

\section{Gravitationally induced particle creation}\label{sect3}

In the present Section, we analyze the physical interpretation of the
curvature-matter coupling in the scalar-tensor representation by adopting
the point of view of the thermodynamics of open systems, in which matter
creation irreversible processes may take place at a cosmological scale \cite%
{Pri0}-\cite{Lima}. As we have already seen in the previous Section, the
energy conservation equation of the curvature-matter coupling, given by Eq.~(%
\ref{eneq}), contains, as compared to the standard adiabatic conservation
equation, an extra term, which can be interpreted in the framework of the
open thermodynamic systems as an irreversible matter creation rate.
According to irreversible thermodynamics, matter creation also represents an
entropy source, generating an entropy flux, and thus leading, in the
presence of the curvature-matter coupling, to a modification in the
temperature evolution.

In the following, we investigate only the case in which all the non-diagonal
components of the energy--momentum tensor of the matter are equal to zero,
so that $T_{\mu \nu}=0$, $\mu \neq \nu $. Generally, the energy-momentum tensor for a viscous dissipative fluid in the presence of heat conduction is given by $T_{\mu \nu}=\left(\rho +p+\Pi\right)-(p+\Pi)g_{\mu \nu}+q_{\mu}u_{\nu}+q_{\nu}u_{\mu}+\pi _{\mu \nu}$, where $\Pi $ is the bulk viscous pressure, $q_{\nu}$ is the heat flux, and $\pi _{\mu \nu}$ is the tensor of viscous dissipation \cite{Mart}. $q_{\mu}$ and $\pi _{\mu\nu}$ must satisfy the conditions $q_{\mu}u^{\mu}=0$ and $\pi _{\mu \nu }u^{\nu}=\pi _{\mu}^{\mu}=0$, respectively. In the following we neglect the viscous effects in the cosmological fluid, thus assuming $\Pi \equiv 0$ and $\pi _{\mu \nu }\equiv 0$. In a comoving reference frame with $u^{\mu}=(1,0)$, the form of the heat flux vector is fixed by the normalization condition as $q^{\mu}=\left(0,\vec{q}\right)$. Therefore in a comoving reference frame all components of the form $u_{\mu }q_{\nu}$ of the energy-momentum tensor are identically equal to zero, and $T_{\mu \nu}$ is a diagonal tensor.
 From the point of view of
the thermodynamics of the irreversible processes, this condition implies the
impossibility of heat transfer in the considered gravitational system. In
particular, this condition is always satisfied in homogeneous and isotropic
cosmological models, described by the Friedmann--Robertson--Walker geometry,
since in these models the condition $T_{0i}\equiv 0$, $i=1,2,3$ must always
hold.

\subsection{Matter creation rates and the creation pressure}

We assume that the cosmological metric is given by the flat isotropic and
homogeneous Friedmann-Robertson-Walker (FRW) metric,
\begin{equation}
ds^2=dt^2-a^2(t)\left(dx^2+dy^2+dz^2\right),
\end{equation}
where $a(t)$ is the scale factor, describing the expansion of the Universe.
In this geometry the cosmological matter is comoving with the cosmological
expansion, and therefore the four velocity of the cosmological fluid is $%
U^{\mu}=(1,0,0,0)$, while the Hubble function takes the form $H=\dot{a}/a$,
since $U^{\mu }\nabla _{\mu }=\dot{}=d/dt$.

To investigate the thermodynamical implications at the cosmological scale
with a curvature-matter coupling we consider that the Universe contains $N$
particles in a volume $V$, with an energy density $\rho $ and a
thermodynamic pressure $p$, respectively. For such a cosmological system,
the second law of thermodynamics, in its most general form, is given by \cite%
{Pri}
\begin{equation}
\frac{d}{dt}\left( \rho a^{3}\right) +p\frac{d}{dt}a^{3}=\frac{dQ}{dt}+\frac{%
\rho +p}{n}\frac{d}{dt}\left( na^{3}\right) ,  \label{21}
\end{equation}%
where $dQ$ is the heat received by the system during time $dt$, and $n=N/V$
is the particle number density, respectively. Due to our choice of the
geometry of the Universe, and of the cosmological principle, only adiabatic
transformations, defined by the condition $dQ=0$, are possible. Therefore in
the following we ignore proper heat transfer processes in the Universe.
However, as one can see from Eq.~(\ref{21}), under the assumption of
adiabatic transformations the second law of thermodynamics contains the term
$[(\rho +p)/n]d\left( na^{3}\right) /dt$, which explicitly takes into
account the time variation of the cosmological particles in a given volume $%
V $. Hence, in the irreversible thermodynamics description of open systems,
even for adiabatic transformations $dQ=0$, one can consider the ``heat''
(internal energy), received/lost by the system, and which is due to the
change in the particle number $n$. For modified gravity with a
curvature-matter coupling the change in the particle number is due to the
transfer of energy from gravity to matter. Thus, via matter creation,
gravity acts as a source of internal energy, and of entropy. For adiabatic
transformations $dQ/dt=0$, we can reformulate Eq.~(\ref{21}) in an
equivalent form as
\begin{equation}
\dot{\rho}+3(\rho +p)H=\frac{\rho +p}{n}\left( \dot{n}+3Hn\right) .
\label{cons0}
\end{equation}

Therefore, from the point of view of the thermodynamics of open systems,
Eq.~(\ref{eneq}), giving the energy balance in the presence of a
curvature-matter coupling, can be interpreted as describing particle
creation in an homogeneous and isotropic geometry, with the time variation
of the particle number density obtained as
\begin{equation}
\dot{n}+3nH=\Gamma n,  \label{22}
\end{equation}%
where the particle creation rate $\Gamma $ is  a non-negative quantity defined as
\begin{equation}  \label{33}
\Gamma =-\frac{1}{\rho +p}\Bigg\{\rho \frac{d}{dt}\ln U(\psi )+\frac{%
2V^{\prime }\left( \psi \right) -U^{\prime }(\psi )L_{m}}{U(\psi )}\dot{\psi}%
\Bigg\}.
\end{equation}
Therefore, the energy conservation equation can be reformulated in the
alternative form
\begin{equation}
\dot{\rho}+3(\rho +p)H=(\rho +p)\Gamma .  \label{41}
\end{equation}

As proven initially in \cite{Pri}, for adiabatic transformations Eq.~(\ref%
{21}), describing irreversible particle creation in an open thermodynamic
system, can be rewritten as an effective energy conservation equation,
\begin{equation}
\frac{d}{dt}\left( \rho a^{3}\right) +\left( p+p_{c}\right) \frac{d}{dt}%
a^{3}=0,
\end{equation}%
or, in an equivalent form, as,
\begin{equation}
\dot{\rho}+3\left( \rho +p+p_{c}\right) H=0,  \label{comp}
\end{equation}%
where we have introduced a new thermodynamic quantity, $p_{c}$, denoted the
creation pressure and defined as \cite{Pri}
\begin{eqnarray}  \label{pc1}
p_{c} &=&-\frac{\rho +p}{n}\frac{d\left( na^{3}\right) }{da^{3}}  \notag \\
&=&-\frac{\rho +p}{3nH}\left( \dot{n}+3nH\right) = -\frac{\rho +p}{3}\frac{%
\Gamma }{H}.
\end{eqnarray}%
Therefore in modified gravity with a linear curvature-matter coupling the
creation pressure is given by
\begin{equation}  \label{pc}
p_{c}=-\frac{1}{3H}\left\{\rho \frac{d}{dt}\ln U(\psi )+\frac{2V^{\prime
}\left( \psi \right) -U^{\prime }(\psi )L_{m}}{U(\psi )}\dot{\psi}\right\}.
\end{equation}

Note that from Eq. (\ref{cons1}), the coupling between the matter and the
higher derivative curvature terms may be interpreted as an exchange of
energy and momentum between both. In the standard formulation of the linear
curvature-matter coupling, by taking into account a FRW background, and from
Eq. (\ref{cons1}) we obtain the energy balance equation as
\begin{equation}
\dot{\rho}+3H(\rho +p)=\frac{\lambda F_{2}(R)}{1+\lambda f_{2}(R)}\,(\alpha
-1)\rho \,\dot{R}.  \label{NMCenergy}
\end{equation}

Hence, by considering that the mechanism for gravitational particle
production is through the nonminimal curvature-matter coupling, so that
comparing Eqs.~(\ref{pc1}) and (\ref{NMCenergy}), we have for the creation
pressure
\begin{equation}
p_{c}=\frac{\lambda F_{2}(R)}{1+\lambda f_{2}(R)}\frac{1}{3H}\,(1-\alpha
)\rho \,\dot{R}\,,  \label{NMCpc}
\end{equation}%
with the requirement that $p_{c}$ be negative. We will only consider the
case of $L_{m}=-p$, i.e., $\alpha =-\omega $, in order to have a
non-vanishing creation pressure.

\subsection{Entropy and temperature evolution}

According to the basic principles of the thermodynamics of open systems the
entropy change consists of two components: the entropy flow term $d_eS$, and
the entropy creation term $d_iS$. The total entropy $S$ of an open
thermodynamic system can be represented as \cite{Pri0,Pri}
\begin{equation}
dS = d_eS + d_iS,
\end{equation}
where by definition $d_iS > 0$. Both the entropy flow and the entropy
production can be obtained from the total differential of the entropy given
by \cite{Pri},
\begin{equation}
T d\left(\bar{s}a^3\right)=d\left(\rho a^3\right)+pda^3-\mu
d\left(na^3\right),
\end{equation}
where $T$ is the temperature of the open thermodynamic system, $\bar{s}%
=S/a^3 $ is the entropy per unit volume, and $\mu $ is the chemical
potential, defined as
\begin{equation}
\mu n=h-T\bar{s},
\end{equation}
where $h=\rho +p$ is the enthalpy of the system.

In the case of a closed thermodynamic system and for adiabatic
transformations we have $dS=0$ and $d_iS=0$. However, in the presence of a
curvature-matter coupling, leading to effective matter creation, there is a
non-zero contribution to the total entropy. For a homogeneous and isotropic
Universe the entropy flow term $d_eS$ vanishes, so that $d_eS = 0$. On the
other hand matter creation also represents a source for entropy creation,
and the time variation of the corresponding entropy is obtained as \cite{Pri}
\begin{eqnarray}  \label{25}
T\frac{d_iS}{dt}&=&T\frac{dS}{dt}=\frac{h}{n}\frac{d}{dt}\left(na^3\right)-%
\mu \frac{d}{dt}\left(na^3\right)  \notag \\
&=&T\frac{\bar{s}}{n}\frac{d}{dt}\left(na^3\right)\geq 0,
\end{eqnarray}

Equation~(\ref{25}) gives the time variation of the entropy as
\begin{equation}
\frac{dS}{dt}=\frac{S}{n}\left( \dot{n}+3Hn\right) =\Gamma S\geq 0,
\label{43}
\end{equation}%
so that the entropy increase due to particle production yields the
expression
\begin{equation}
S(t)=S_{0}e^{\int_{0}^{t}{\Gamma \left( t^{\prime }\right) dt^{\prime }}},
\end{equation}
where $S_{0}=S(0)$ is a constant. With the use of Eq.~(\ref{43}), we obtain
for the entropy creation in the scalar-tensor representation of the linear
coupling between matter and geometry the following equation
\begin{equation}
\frac{1}{S}\frac{dS}{dt}=-\frac{1}{\rho +p}\Bigg\{\rho \frac{d}{dt}\ln
U(\psi )+\frac{2V^{\prime }\left( \psi \right) -U^{\prime }(\psi )L_{m}}{%
U(\psi )}\dot{\psi}\Bigg\}.  \label{entff}
\end{equation}

The entropy flux four-vector $S^{\mu }$ is defined as \cite{Cal}
\begin{equation}
S^{\mu }=n\sigma U^{\mu },
\end{equation}%
where $\sigma =S/N$ is the specific entropy per particle. $S^{\mu }$ must
satisfy the second law of thermodynamics, which imposes the constraint $%
\nabla _{\mu }S^{\mu }\geq 0$ on its four-dimensional divergence. The Gibbs
relation \cite{Cal},
\begin{equation}
nTd\sigma =d\rho -\frac{h}{n}dn,
\end{equation}%
together with the definition of the chemical potential $\mu $ of the open
thermodynamic system,
\begin{equation}
\mu =\frac{h}{n}-T\sigma ,
\end{equation}%
yields
\begin{eqnarray}
\nabla _{\mu }S^{\mu } &=&\left( \dot{n}+3nH\right) \sigma +nU^{\mu }\nabla
_{\mu }\sigma  \notag  \label{48} \\
&=&\frac{1}{T}\left( \dot{n}+3Hn\right) \left( \frac{h}{n}-\mu \right) ,
\end{eqnarray}%
where we have used the relation
\begin{equation}
nT\dot{\sigma}=\dot{\rho}-\frac{\rho +p}{n}\dot{n}=0,
\end{equation}%
which immediately follows from Eq.~(\ref{cons0}). With the use of Eq.~(\ref%
{22}) we obtain the entropy production rate due to the particle creation
processes given by
\begin{eqnarray}
\nabla _{\mu }S^{\mu }&=&\Gamma \frac{n}{T}\left( \frac{h}{n}-\mu \right) =-%
\frac{n}{T(\rho +p)}\Bigg\{\rho \frac{d}{dt}\ln U(\psi )  \notag \\
&&+\frac{2V^{\prime }\left( \psi \right) -U^{\prime }(\psi )L_{m}}{U(\psi )}%
\dot{\psi}\Bigg\}\left( \frac{h}{n}-\mu \right) .
\end{eqnarray}

A general thermodynamic system is described by two fundamental thermodynamic
variables, the particle number density $n$, and the temperature $T$,
respectively. If the system is in an equilibrium state, the energy density $%
\rho $ and the thermodynamic pressure $p$ are obtained, in terms of $n$ and $%
T$, from the equilibrium equations of state of the matter,
\begin{equation}
\rho =\rho (n,T), \qquad   p=p(n,T).  \label{51}
\end{equation}%
Therefore the energy conservation equation (\ref{41}) can be obtained in
the following general form
\begin{equation}
\frac{\partial \rho }{\partial n}\dot{n}+\frac{\partial \rho }{\partial T}%
\dot{T}+3(\rho +p)H=\Gamma n.
\end{equation}%
By using the general thermodynamic relation \cite{Cal}
\begin{equation}
\frac{\partial \rho }{\partial n}=\frac{h}{n}-\frac{T}{n}\frac{\partial p}{%
\partial T},
\end{equation}%
it follows that the temperature evolution of the newly created particles due
to the curvature-matter coupling is given by the expression
\begin{equation}
\frac{\dot{T}}{T}=c_{s}^{2}\frac{\dot{n}}{n}=c_{s}^{2}\left( \Gamma
-3H\right) ,  \label{54}
\end{equation}%
where the speed of sound $c_{s}$ is defined as $c_{s}^{2}=\partial
p/\partial \rho $. If the geometrically created matter satifies a barotropic
equation of state of the form $p=\left( \gamma -1\right) \rho $, $1\leq
\gamma \leq 2$, the temperature evolution follows the simple equation
\begin{equation}
T=T_{0}n^{\gamma -1}.
\end{equation}

\subsection{Bulk-viscosity description of matter creation processes with a curvature-matter coupling}

An alternative physical interpretation of particle creation processes in
cosmology was suggested by Zeldovich \cite{Zeld}, and later on by Murphy
\cite{Mur} and Hu \cite{Hu}. According to this interpretation, the viscosity
of the cosmological fluid represents a phenomenological description of the
effect of the creation of particles by the non-stationary gravitational
field of the expanding universe. Therefore, from a physical point of view, a
non-vanishing particle production rate is equivalent to the introduction of
a bulk viscous pressure in the energy-momentum tensor of the cosmological
fluid. From a quantum mechanical point of view, such a viscous pressure can
also be related to the viscosity of the vacuum \cite{Zeld,Mur,Hu}. This
physical interpretation follows from the simple circumstance that any source
term in the energy balance equation of a general relativistic fluid may be
formally rewritten in terms of an effective bulk viscosity \cite{Mart}.

The energy-momentum tensor of a general relativistic fluid with bulk
viscosity as the only dissipative process can be written as \cite{Mart}
\begin{equation}
T_{\mu \nu}=\left(\rho +p+\Pi \right)U_{\mu}U_{\nu}-\left(p+\Pi\right) g_{\mu\nu},
\end{equation}
where $\Pi $ is the bulk viscous pressure. The particle flow vector $N^{\mu}$
is defined as $N^{\mu}=nU^{\mu}$. In the framework of causal thermodynamics
the entropy flow vector $S^{\mu}$ takes the form \cite{Isr}
\begin{equation}  \label{flow}
S^{\mu}=sN^{\mu}-\frac{\tau \Pi^2}{2\xi {\mathcal{T}}}U^{\mu},
\end{equation}
where $\tau $ is the relaxation time, and $\xi $ is the coefficient of bulk
viscosity. In Eq.~(\ref{flow}), we have limited ourselves to considering only
second-order deviations from equilibrium. In the case of homogeneous and
isotropic geometries, in the presence of bulk viscous dissipative phenomena,
the energy conservation equation is obtained as
\begin{equation}  \label{comp1}
\dot{\rho}+3\left(\rho +p+\Pi\right)H=0.
\end{equation}

By comparing Eq.~(\ref{comp1}), giving the energy conservation equation for
a cosmological fluid in the presence of bulk viscosity, with Eq.~(\ref{comp}),
which includes in the energy balance the creation of particles from the
gravitational field due to the curvature-matter coupling, it follows that
these two equations are equivalent if
\begin{equation}
p_c=\Pi=-\frac{1}{3H}\Bigg\{\rho \frac{d}{dt}\ln U(\psi )+\frac{2V^{\prime
}\left( \psi \right) -U^{\prime }(\psi )L_{m}}{U(\psi )}\dot{\psi}\Bigg\}.
\end{equation}
Therefore particle creation can be indeed described from a phenomenological
point of view by introducing an effective bulk viscous pressure in the
energy-momentum tensor of the cosmological fluid. Hence it follows that the
causal bulk viscous pressure $\Pi $ acts as a creation pressure.

Hence, it would be interesting to investigate matter creation processes in
modified gravity with a curvature-matter coupling from the point of view of
bulk viscous thermodynamic processes. Hence we shall consider in the
following that there is a change in the number of particles, due to matter
creation processes, with bulk viscous pressure playing the role of the
creation pressure. We introduce a simple toy model in which the newly
created particles obey, as a function of the particle number density $n$, an
equation of state of the form
\begin{equation}
\rho =\rho _{0}\left( \frac{n}{n_{0}}\right) ^{\gamma }=kn^{\gamma }, \qquad
p=(\gamma -1)\rho ,  \label{9_1}
\end{equation}%
where $\rho _{0}$, $n_{0}$ and $\gamma $ are constants, we have denoted $%
k=\rho _{0}/n_{0}^{\gamma }$, and $1\leq \gamma \leq 2$, respectively. Using
Eq.~(\ref{9_1}), then Eq.~(\ref{comp1}) takes the form of a particle balance
equation,
\begin{equation}
\dot{n}+3Hn=\Gamma n,
\end{equation}%
where
\begin{equation}
\Gamma =-\frac{\Pi }{\gamma H}
\end{equation}%
is the particle production rate, proportional to the bulk viscous pressure.
Combining the equation of state Eq.~(\ref{9_1}) with the Gibbs relation $Td%
\bar{s}=d(\rho /n)+pd(1/n)$ we obtain $\bar{s}=\bar{s_{0}}=\mathrm{constant}$,
that is, particles are created with constant entropy density.

However, there is a major difference between the particle creation
irreversible processes in open thermodynamic systems and bulk viscous
processes, and this difference is related to the expression for entropy
production rate. While the entropy production rate associated to particle
creation is given by \cite{Cal}
\begin{equation}
\nabla _{\mu }S^{\mu }=-\frac{3Hp_c}{{T}}\left(1+\frac{\mu \Gamma n}{3Hp_c}%
\right)\geq 0,
\end{equation}
in the presence of bulk viscous dissipative processes the entropy production
rate can be obtained as \cite{Mart}
\begin{equation}
\nabla _{\mu}S^{\mu}=-\frac{\Pi}{{T}}\Bigg[3H+\frac{\tau }{\xi}\dot{\Pi}+%
\frac{\tau }{2\xi }\Pi\left(3H+\frac{\dot{\tau}}{\tau }-\frac{\dot {\xi}}{\xi%
}-\frac{\dot{{T}}}{{T}}\right)\Bigg].
\end{equation}

In the particle creation model in open thermodynamics systems the entropy
production rate is proportional to the creation pressure, while in the
viscous dissipative processes thermodynamic interpretation $\nabla _{\mu
}S^{\mu }$ is quadratic in the creation pressure, $\nabla _{\mu }S^{\mu
}\propto p_{c}^{2}/\xi {T}$, and, moreover, involves a new dynamical
variable, the bulk viscosity coefficient.

\section{Cosmological applications}\label{sect4}

In the present Section, we consider several cosmological applications of the
scalar-tensor formulation of modified gravity with a linear curvature-matter
coupling, as its interpretation as a particle creation theory. For a
homogeneous and isotropic geometry the gravitational field equations (\ref{15})-(\ref{eqff}) take the form
\begin{eqnarray}  \label{eqex1}
3H^{2} &=&\frac{\ddot{\psi}}{\psi }+\left( \frac{2}{3}\rho +p\right) \frac{%
U(\psi )}{\psi }  \notag \\
&&-\frac{1}{3}\left[ \frac{V(\psi )}{\psi }+2V^{\prime }(\psi )-2U^{\prime
}(\psi )L_{m}\right] ,
\end{eqnarray}%
\begin{eqnarray}  \label{eqex2}
2\dot{H}+3H^{2} &=&-\frac{\dot{a}}{a}\frac{\dot{\psi}}{\psi }-\frac{1}{3}%
\rho \frac{U\left( \psi \right) }{\psi }  \notag \\
&&-\frac{1}{3}\left[\frac{V(\psi )}{\psi }+2V^{\prime }(\psi )-2U^{\prime
}(\psi )L_{m}\right],
\end{eqnarray}%
and
\begin{equation}
-3\left( \dot{H}+2H^{2}\right) -V^{\prime }\left( \psi \right) +U^{\prime
}\left( \psi \right) L_{m}=0,  \label{eqex3}
\end{equation}
respectively.

Equation~(\ref{22}) can be integrated to give the particle number time
variation as
\begin{equation}
n(t)=\frac{n_{0}}{a^{3}}e^{\int \Gamma \left( t\right) dt},
\end{equation}%
where $n_{0}$ is an arbitrary constant of integration, while, by assuming a
barotropic equation of state of the form $p=p\left( \rho \right) $ we obtain
for the density evolution
\begin{equation}  \label{69}
\int \frac{d\rho }{\rho +p\left( \rho \right) }=-3\ln a+\int {\Gamma (t)dt}%
+\ln \rho _{0},
\end{equation}%
where $\rho _{0}$ is an integration constant.

\subsection{Cosmological models satisfying the condition $2V^{\prime }(%
\protect\psi )-U^{\prime }(\protect\psi )L_{m}=0$.}

The linear curvature-matter coupling depends on the two arbitrary (and independent) potentials $U(\psi )$ and $V(\psi )$.
As a simple toy model, in the following, we assume that the two potentials $%
U $ and $V$ are related to the matter Lagrangian via the relation
\begin{equation}
2V^{\prime }(\psi )-U^{\prime }(\psi )L_{m}=0.  \label{as1}
\end{equation}%
Moreover, we restrict our analysis to the case of dust, with negligible
thermodynamic pressure $p=0$. With the choice of Eq. (\ref{as1}), the particle
creation rate, given by Eq.~(\ref{33}), takes the form
\begin{equation}  \label{as2}
\Gamma =-\frac{d}{dt}\ln U(\psi (t))=-\frac{U^{\prime }(\psi)}{U(\psi)}\dot{%
\psi }\geq 0.
\end{equation}
Therefore, in this approach the matter creation rate is determined by the
function $U(\psi )$, which describes the coupling between the matter
Lagrangian and the scalar field, its derivative, and the time variation of
the scalar field $\psi$ only.

With this choice, the variations of the particle number and of the matter
energy density is given by
\begin{equation}  \label{au1}
n(t)=\frac{n_{0}}{a^{3}U(\psi )}, \qquad \rho (t)=\frac{\rho _{0}}{%
a^{3}U(\psi )}.
\end{equation}

With the assumption of Eq. (\ref{as1}) on the potentials, the gravitational field
equations (\ref{eqex1})-(\ref{eqex3}) take the form
\begin{equation}  \label{exf1}
3H^{2}=\frac{\ddot{\psi}}{\psi }+\frac{2}{3}\rho \frac{U(\psi )}{\psi }-%
\frac{1}{3}\left[ \frac{V(\psi )}{\psi }-U^{\prime }(\psi )L_{m}\right] ,
\end{equation}
\begin{equation}  \label{exf2}
2\dot{H}+3H^{2}=-\frac{\dot{a}}{a}\frac{\dot{\psi}}{\psi }-\frac{1}{3}\rho
\frac{U\left( \psi \right) }{\psi }-\frac{1}{3}\left[\frac{V(\psi )}{\psi }%
-U^{\prime }(\psi )L_{m}\right],
\end{equation}
\begin{equation}  \label{exf3}
6\left( \dot{H}+2H^{2}\right) =U^{\prime }\left( \psi \right) L_{m},
\end{equation}
respectively.

In the following, we consider only de Sitter type accelerating solutions of
the system given by Eqs.~(\ref{exf1})-(\ref{exf3}), with $H=H_{0}=\mathrm{constant}$
and $a(t)=\exp \left( H_{0}t\right) $. Then we obtain first
\begin{equation}
U^{\prime }\left( \psi \right) L_{m}=12H_{0}^{2},  \label{eqUf}
\end{equation}%
while the evolution equation for the scalar field $\psi $, which can be
obtained from Eqs.~(\ref{au1})-(\ref{exf2}), is given by
\begin{equation}
\ddot{\psi}+H_{0}\dot{\psi}+\frac{\rho _{0}}{e^{3H_{0}t}}=0,
\end{equation}%
with the general solution given by
\begin{eqnarray}  \label{79}
\psi (t)&=&-\frac{1}{6H_{0}^{2}}\Bigg[ \rho _{0}e^{-3H_{0}t}-3 \rho
_{0}e^{-H_{0}(t+2t_{0})}+2\rho _{0}e^{-3H_{0}t_{0}}  \notag \\
&&+6H_{0}\psi _{01}e^{-H_{0}(t-t_{0})}-6H_{0}(H_{0}\psi _{0}+ \psi _{01})%
\Big] ,
\end{eqnarray}
where we have used the initial conditions $\psi \left( t_{0}\right) =\psi
_{0}$, and $\psi ^{\prime }\left( t_{0}\right) =\psi _{01}$, respectively.
Since for the dust fluid the matter Lagrangian is $L_{m}=\rho $, Eq. ($\ref%
{eqUf})$ yields
\begin{equation}
\frac{U^{\prime }\left( \psi \right) }{U\left( \psi \right) }=12\frac{%
H_{0}^{2}}{\rho _{0}}e^{3H_{0}t},
\end{equation}
or, equivalently,
\begin{equation}  \label{80}
\frac{1}{U(t)}\frac{dU(t)}{dt}=12\frac{H_0^2}{\rho _0}e^{3H_0t}\dot{\psi}.
\end{equation}

The above equation determines the time variation of the potential $U(\psi )$
as
\begin{equation}\label{Uimp}
U\left( t \right) =U_{0}\exp \left[\frac{6 H_0 \psi_{01} e^{H_0 (2 t+t_0)}}{%
\rho _0}-3 e^{2 H_0 (t-t_0)}+6 H_0 t\right],
\end{equation}
where $U_0$ is an arbitrary constant of integration. The time variation of
the particle creation rate is obtained as
\begin{equation}
\Gamma (t)=\frac{6H_0U_0}{\rho _0}\left[2H_0|\psi
_{01}|e^{H_0\left(2t+t_0\right)}+\rho _0e^{2H_0\left(t-t_0\right)}-\rho _0%
\right],
\end{equation}
where we have assumed $\psi _{01}<0$. The particle creation rate is
non-negative for all times $t$, and is a monotonically increasing function of
time for all $t\geq t_0$, and its initial value is given by $\Gamma
\left(t_0\right)=12H_0^2|\psi _{01}|/\rho _0\geq 0$.

Therefore we obtain for the time variation of the matter energy density the
equation
\begin{equation}
\rho \left( t\right) =\frac{\rho _0}{U_0}\exp \left[\frac{3 e^{2 H_0
(t-t_0)} \left(2 H_0 |\psi _{01}| e^{3 H_0 t_0}+\rho _0\right)}{\rho _0}-9 H_0
t\right].
\end{equation}
For the time variation of the potential $V\left(t \right) $ we obtain the
evolution equation
\begin{equation}
2\frac{dV}{dt}=\rho \frac{dU}{dt},
\end{equation}
which gives the potential $V$ in an integral form as
\begin{eqnarray}
V(t)=V_0+3 H_0 U_0 \int e^{H_0 \left[\frac{12 |\psi _{01}| e^{H_0 (2 t+t_0)}%
}{\rho _0}-3 t\right]} \times  \notag \\
\times \left[-\rho _0 e^{2 H_0 (t-t_0)}+ 2 H_0 |\psi _{01}| e^{H_0 (2 t+t_0)}+\rho
_0\right] dt,
\end{eqnarray}
where $V_{0}$ is an arbitrary constant of integration. The creation pressure, defined as $p_{c}=-\rho\Gamma /(3H)$, is given by
\begin{eqnarray}
p_{c}&=&2 U_0 \left[-\rho _0 e^{2 H_0
(t-t_0)}-2 H_0 |\psi _{01}| e^{H_0 (2 t+t_0)}+ \rho _0\right] \times \notag \\
&& \hspace{-0.75cm} \times \exp \left[\frac{3 e^{2 H_0 (t-t_0)} \left(2 H_0 |\psi _{01}|
e^{3 H_0 t_0}+\rho _0\right)}{\rho _0}- 9 H_0 t\right].
\end{eqnarray}

Finally, for the comoving entropy of the de Sitter type expanding Universe
in the linear curvature-matter coupling theory we obtain
\begin{eqnarray}
S(t)&=&\frac{S_{0}}{U\left( t \right) }=\frac{S_{0}}{U_{0}}\exp \Big[ %
3e^{2H_{0}(t-t_{0})} - 6H_{0}t  \notag \\
&&+\frac{6H_{0}|\psi _{01}|e^{H_{0}(2t+t_{0})}}{\rho _{0}} \Bigg] .
\end{eqnarray}

The entropy is a monotonically increasing function of time, with the
property $\dot{S}(t)\geq 0$, $\forall t\geq t_0$. In the first order approximation, and for small times we obtain for the entropy the following expression
\be
S(t)\propto \exp\left(\frac{12H_0^2|\psi _{01}|}{\rho _0}t\right).
\ee

Therefore, the curvature-matter
coupling allows the production of a large amount of entropy during a de
Sitter type evolutionary phase of the Universe.

\subsection{de Sitter type expansionary models with constant matter creation
rate}

In the following, in the scalar-tensor representation of $Lf\left(R,L_m\right)$ gravity, we consider a second simple cosmological toy model by assuming that the cosmological expansion of a dust Universe, with $p=0$, is accelerating with $a=\exp\left(H_0t\right)$, where $H_0=\mathrm{constant}$, and the particle creation rate is a constant during
the entire accelerating phase, and it is given by $\Gamma =\Gamma _0=3H_0=%
\mathrm{constant}$. Moreover, we take the matter Lagrangian as $L_m=\rho$.
Then from Eq.~(\ref{69}) it follows immediately that the matter density of
the Universe is also a constant,
\begin{equation}
\rho =\rho _0=\mathrm{constant}.
\end{equation}
Then the constancy of the matter creation rate, given by Eq.~(\ref{33})
imposes the following condition on the potentials $U$ and $V$,
\begin{equation}  \label{89}
2\dot{V}+3H_0\rho _0U=0,
\end{equation}
while the field equation Eq.~(\ref{eqex3}) yields
\begin{equation}  \label{90}
\dot{V}-\rho _0\dot{U}=-6H_0^2\dot{\psi}.
\end{equation}

From the field equations Eqs.~(\ref{eqex1}) and (\ref{eqex2}) we obtain the
evolution equation for $\psi $ as
\begin{equation}
\ddot{\psi}+H_{0}\dot{\psi} +\rho _{0}U(\psi (t) )=0.  \label{91}
\end{equation}
From Eqs.~(\ref{89}) and (\ref{90}) we obtain
\begin{equation}
\frac{3H_{0}\rho _{0}}{2}U+\rho _{0}\dot{U}=6H_{0}^{2}\dot{\psi}.
\end{equation}%

By taking the time derivative of the above equation, and by eliminating $%
\ddot{\psi}$ with the help of Eq.~(\ref{91}), it follows that $U$ satisfies
the equation
\begin{equation}
\ddot{U}+\frac{5}{2}H_{0}\dot{U}+\frac{15}{2}H_{0}^{2}U=0,
\end{equation}%
with the general solution given by
\begin{eqnarray}  \label{Uf}
U(t) &=&\frac{e^{\frac{5}{4}H_{0}(t_{0}-t)}}{215H_{0}}\Bigg\{
215H_{0}U_{0}\cos \left[ \frac{1}{4}\sqrt{215}H_{0}(t-t_{0})\right]
   \nonumber \\
&& \hspace{-0.75cm}  \sqrt{215}(5H_{0}U_{0}+4U_{01})
\sin \left[ \frac{1}{4}\sqrt{215}H_{0}(t-t_{0})\right] \Bigg\},
\end{eqnarray}%
where we have used the initial conditions $U\left( t_{0}\right) =U_{0}$ and $%
U^{\prime }\left( t_{0}\right) =U_{01}$, respectively. The time dependence
of the potential $V$ is obtained from Eq.~(\ref{89}) in the form
\begin{widetext}
\begin{eqnarray}
V(t)=\frac{\rho _{0}e^{\frac{5}{4}H_{0}(t_{0}-t)}}{860H_{0}}\left\{ \sqrt{%
215}(2U_{01}-19H_{0}U_{0})\sin \left[ \frac{1}{4}\sqrt{215}H_{0}(t-t_{0})\right]
+43(5H_{0}U_{0}+2U_{01})\cos \left[ \frac{1}{4}\sqrt{215}H_{0}(t-t_{0})\right] \right\} .
\end{eqnarray}

With the use of the explicit time dependence of $U$ in Eq.~(\ref{91}), we
obtain for $\psi (t)$ the following expression
\begin{eqnarray}
\psi (t) &=&\frac{1}{11610H_{0}^{3}}\Big\{-430e^{H_{0}(t_{0}-t)}\left(%
27H_{0}^{2}\psi _{01}- 3H_{0}\rho _{0}U_{0}-2\rho _{0}U_{01}\right)+387\left(30H_{0}^{3}\psi _{0}+30H_{0}^{2}\psi _{01}-5H_{0}\rho _{0}U_{0}-2\rho _{0}U_{01}\right)
\notag \\
&& \hspace{-1.25cm}
+ \rho _{0}e^{\frac{5}{4}H_{0}(t_{0}-t)}\left[\sqrt{215}%
(39H_{0}U_{0}+14U_{01}) \sin \left( \frac{1}{4}\sqrt{215}H_{0}(t-t_{0})\right)
+43(15H_{0}U_{0}-2U_{01})\cos \left( \frac{1}{4}\sqrt{215}H_{0}(t-t_{0})\right) \right]\Bigg\},
\end{eqnarray}%
\end{widetext}
where we have used the initial conditions $\psi \left( t_{0}\right) =\psi
_{0}$ and $\dot{\psi}\left( t_{0}\right) =\psi _{01}$, respectively.

Due to the cosmological particle production the  entropy of the
Universe increases as
\begin{equation}
S(t)=S_0e^{\int{\Gamma (t)dt}}=S_0e^{3H_0t}.
\end{equation}
However, the specific entropy $s=S/V$ remains a constant during the
cosmological evolution, $s=s_0=\mathrm{constant}$.

\section{Total entropy behavior in $Lf\left(R,L_m\right)$ gravity with particle creation}\label{sectn}

In the present paper, we have defined the entropy through the particle
production rate, given by Eq.~(43), as depending on the positive particle
creation rate $\Gamma $ via the relation
\begin{equation}
\frac{\dot{S}}{S}=\Gamma \geq 0.
\end{equation}
Therefore, in an ever expanding Universe with particle creation, the matter
entropy will increase indefinitely. On the other hand, all natural systems
tend to approach a state of thermodynamic equilibrium, implying that the
entropy of equilibrium systems never decreases, $\dot{S}\geq 0$, and that it is
concave when approaching the equilibrium state, $\ddot{S}\leq 0$. However,
in the present, and several other, cosmological models, these fundamental
requirements for the behavior of the entropy do not seem to be satisfied.

The problem of the validity of the second law of thermodynamics in cosmology was
investigated in detail in \cite{Mim,Pav}, where it was shown that
the Universe approaches thermodynamic equilibrium in a de Sitter phase,
if one defines the total entropy $S_{tot}$ of the Universe as the entropy of
the apparent horizon plus that of matter and radiation inside it. Then it
follows that $S_{tot}$ increases, and that it is concave, thus leading to
the result that the second law of thermodynamics is still valid for the case
of the cosmological expansion. In the following, we investigate the
thermodynamic properties of the total entropy in a Universe with matter
creation.

In the standard thermodynamic description of physical systems the time
parameter $t$ is not a thermodynamic equilibrium variable. Therefore, the
variation of the thermodynamic quantities should be considered with respect
to some extensive variable. In the following we will adopt, following \cite%
{Mim,Pav}, as extensive variable for the cosmological system the proper volume enclosed by the apparent horizon, or, more specifically, its scale factor $a$. Then the
relation between $d/dt$ and $d/da$ is simply
\begin{equation}
\frac{d}{dt} =a H \frac{d}{da}.
\end{equation}
In the following we denote by a prime the derivative with respect to the
extensive variable $a$.

We define the total entropy of a FRW Universe with dust as the sum of the
entropy of the apparent horizon $S_{ah}$, proportional to its area, and that
of the matter particles within it $S_m$ \cite{Mim,Pav}. For practical
purposes, in the case of the flat FRW model, the total entropy is
\begin{equation}
S_{tot}= S_{ah} + S_m = \frac{\pi}{H^2} + \frac{4\pi}{3H^3}\,n(t)\; ,
\label{MP}
\end{equation}
where we have used the fact that the radius of the apparent horizon is $%
r_{ah}= H^{-1}$ \cite{Bak}. We also introduce an important observational quantity, the
deceleration parameter $q$, defined as
\begin{equation}
q=\frac{d}{dt}\frac{1}{H}-1=-a\frac{H^{\prime }}{H}-1.
\end{equation}

Therefore, the variation of the total entropy can be obtained as
\begin{equation}
\frac{ S_{tot}^{\prime }}{S_{tot}} = \frac{S_{ah}^{\prime
}+S_m^{\prime}}{S_{ah} + S_m},
\end{equation}
\begin{equation}
\frac{ S_{tot}^{\prime \prime }}{S_{tot}} = \frac{S_{ah}^{\prime \prime
}+S_m^{\prime \prime }}{S_{ah} + S_m},
\end{equation}
respectively.

The particle number $n$ satisfies Eq.~(\ref{22}), and is rewritten as
\begin{equation}
aH\frac{dn}{da}+3nH=\Gamma n.
\end{equation}%
From its definition the derivatives of the entropy with respect to the scale
factor can be evaluated as
\begin{equation}
S_{tot}^{\prime}=-\frac{2\pi a}{H^{2}(a)}H^{\prime }(a)+\frac{4\pi n(a)}{%
H^{3}(a)}\left[ \frac{\Gamma (a)}{3}-aH^{\prime }(a)-H(a)\right] ,
\end{equation}%
\begin{eqnarray}
S_{tot}^{\prime \prime }&=&\frac{2\pi }{3aH^{4}(a)}\Big\{ 2H(a)\Big[ %
3a^{2}H^{\prime 2}(a)  \notag \\
&&+n(a)\left[ a\left[ \Gamma ^{\prime }(a)- 3aH^{\prime\prime }(a)+12H^{\prime
}(a)\right] -6\Gamma (a)\right] \Big]   \notag \\
&&-3H^{2}(a)\Big[ a\left[ aH^{\prime \prime }(a)+H^{\prime }(a)\right] -6n(a)%
\Big] \notag \\
&&+2n(a)\left[ \Gamma (a)-3aH^{\prime }(a)\right]^{2}  \Big\}  ,
\end{eqnarray}
respectively.

In terms of the deceleration parameter we can express the variation of the
total entropy with respect to $a$ as
\begin{equation}
S_{tot}^{\prime }=\frac{2\pi }{H}\left( q+1\right) +\frac{4\pi }{H^{3}}%
\left( qH+\frac{\Gamma }{3}\right) n,
\end{equation}%
\begin{eqnarray}
&&S_{tot}^{\prime \prime }=\frac{2 \pi } {3 H^3(a)}\Big\{2 \Gamma (a)
n(a) \left[\Gamma (a)-3 a H^{\prime }(a)\right] \notag \\
&& + 2 H(a) n(a) \Big[a \left[\Gamma ^{\prime }(a)-6 q(a) H^{\prime
}(a)\right] 3 \Gamma (a) (q(a)-1)\Big]  \notag \\
&& -3 H^2(a) \Big[a [q(a)+1] H^{\prime }(a)+n(a) \left[6 q(a)-2 a q^{\prime
}(a)\right]\Big]  \notag \\
&& + 3 a H(a)^3 q^{\prime }(a)\Big\}.
\end{eqnarray}

Therefore the standard thermodynamic requirements $S_{tot}^{\prime}\geq
0 $ and $S_{tot}^{\prime \prime }\leq 0$ impose the following constraints
on the particle creation rate $\Gamma $, and its derivative with respect to
the scale factor
\begin{equation}
\Gamma (a)\geq \frac{3a\left[ H(a)+2n(a)\right] H^{\prime }(a)}{2n(a)}+3H(a),
\end{equation}%
\begin{equation}
\Gamma (a)\geq -\frac{3\left[ q(a)+1\right] H^{2}(a)}{2n(a)}-q(a)H(a),
\end{equation}%
and
\begin{widetext}
\begin{eqnarray}
\Gamma ^{\prime }(a) & \leq & \frac{1}{2aH(a)n(a)}\Big\{ 6H(a)\Big[ n(a)%
\left[ a^{2}H^{\prime \prime }(a)+2\Gamma (a)-
4aH^{\prime }(a)\right] -a^{2}H^{\prime 2}(a)\Big] -2n(a)\left[ \Gamma
(a)- 3aH^{\prime }(a)\right] ^{2}  \notag \\
&& + 3H^2(a)\left[ a\left( aH^{\prime \prime }(a)+H^{\prime }(a)\right) -6n(a)%
\right] \Big\} ,
\end{eqnarray}
\begin{eqnarray}
\Gamma ^{^{\prime }}(a)& \leq & \frac{1}{2 a H(a) n(a)}\Big\{-2 \Gamma (a)
n(a) \left[\Gamma (a)-3 a H^{\prime }(a)\right]+
3 H^2(a) \left[a (q(a)+1) H^{\prime }(a)+n(a) \left(6 q(a)-2 a q^{\prime
}(a)\right)\right]+  \notag \\
&&6 H(a) n(a) \left[\Gamma (a)+2 a q(a) H^{\prime }(a)+\Gamma (a) (-q(a))%
\right]-
3 a H^3(a) q^{\prime }(a)\Big\},
\end{eqnarray}
\end{widetext}
respectively.
Since in the $Lf\left( R,L_{m}\right) $ gravity theory the matter creation
rate $\Gamma $ is determined by the coupling functions $U$ and $V$, the
thermodynamic conditions impose some strong constraints on the allowed
physical form of these functions.

A particularly interesting case is that of the de Sitter evolution of
the Universe, with $H=H_{0}=\mathrm{constant}$. For this situation the total
entropy of the Universe is given by
\begin{equation}
S_{tot}=\frac{\pi }{H_{0}^{2}}+\frac{4\pi }{3H_{0}^{3}}n(a),
\end{equation}%
giving
\begin{equation}
S_{tot}^{\prime }=\frac{4\pi }{3H_{0}^{3}}n^{^{\prime }}(a)=\frac{4\pi }{%
3H_{0}^{4}}\frac{\left[ \Gamma (a) -3H_{0} \right]}{a}n(a) \geq 0,
\end{equation}%
\begin{eqnarray}
S_{tot}^{\prime \prime }&=&\frac{4 \pi n(a)}{3 a^2 H_0^5} \Big\{\Gamma
^2(a)+H_0 \left[a \Gamma ^{\prime }(a)+12 H_0\right]  \notag \\
&&-7 H_0 \Gamma (a)\Big\} \leq 0.
\end{eqnarray}

The thermodynamic condition of the non-negativity of the total entropy
derivative with respect to the scale factor imposes the conditions $\Gamma
\geq 3H_0$ and $\Gamma ^{\prime }(a)\leq \left[7\Gamma (a)-\Gamma
^2(a)/H_0-12H_0\right]/a$ on the particle creation rate $\Gamma$. In the
particular case $\Gamma =3H_0$, we obtain $S_{tot}=\mathrm{constant}$,
showing that in this case the cosmological evolution is isentropic, with the
total entropy being a constant.

\section{Discussion and conclusions}\label{sect5}

In the present paper, we have considered the thermodynamic interpretation of
modified theories of gravity with a linear coupling between matter and
geometry, which we denote as $Lf\left(R,L_m\right)$ gravity. This theory
represents a particular class, corresponding to a specific choice of the
gravitational Lagrangian, of a very general class of theories, in which the
action is an arbitrary function of the Ricci scalar and of the matter
Lagrangian. An interesting characteristic of these theories is the
non-conservation of the energy-momentum tensor of the matter, indicating
that matter and energy fluxes can be generated by the conversion of
the geometric curvature, describing the gravitational field,  into matter. Hence the presence of matter, and its possible
coupling to geometry could modify the cosmological evolution in a way that
goes far beyond the standard description of general relativity. The presence
of a source term in the energy balance equation can be naturally interpreted
in the framework of the thermodynamics of open systems as describing a
particle creation process, in which the ``geometric energy'' of the
gravitational field is transferred to ``real'' matter. During the particle
production phase a large amount of entropy is produced.

In order to estimate the effective thermodynamics quantities we have first
introduced the equivalent scalar-tensor representation of the $%
Lf\left(R,L_m\right)$ theory, which can be formulated in terms of a scalar
field $\psi $, with two independent potentials $V(\psi )$ and $U(\psi )$,
with the potential $U(\psi )$ coupled to the matter Lagrangian. Using the
scalar-tensor representation of the $Lf\left(R,L_m\right)$ theory, we have
obtained the particle creation rate, the creation pressure and the entropy
associated to the gravitational energy transfer to matter. The cosmological
implications of the particle creation have also been investigated, by
assuming a specific relation between the two potentials. The imposed
condition makes the particle creation rate a function of the second scalar
potential $U(\psi)$, which directly couples to the matter Lagrangian. The
gravitational field equations corresponding to these choices have a de
Sitter type accelerating solution, where the cosmic acceleration is triggered by
the particle creation process, which generates a negative creation pressure.
Thus, it was argued that the
negative creation pressure is responsible for the accelerated expansion of
the Universe.

Matter creation processes are supposed to play a fundamental role in the quantum field theoretical approaches to gravity, where they naturally appear. It is a standard result of quantum field theory in curved spacetimes that quanta of the minimally-coupled scalar field are created in the expanding Friedmann-Robertson-Walker universe \cite{Parker}. That's why finding an equivalent microscopic quantum description of the matter creation processes considered in the present paper could shed some light on the physical mechanisms leading to particle generation via gravity and matter geometry coupling. In the following we will briefly point out that such mechanisms do exist, and can be understood, at least qualitatively, in the framework of some semiclassical gravity models.

In semiclassical gravity it is assumed that the gravitational field remains classical, while the classical bosonic fields $\phi $ are quantized. In order to couple quantized fields to classical gravitational fields the quantum energy momentum tensor $\hat{T}_{\mu \nu}$ is replaced by its expectation value  with respect to some quantum state $\Psi $, thus leading to the effective semiclassical Einstein equation \cite{Carl},
\be\label{gr1}
R_{\mu \nu}-\frac{1}{2}g_{\mu \nu}R=\frac{8\pi G}{c^4}\left<\Psi \right |\hat{T}_{\mu \nu}\left |\Psi \right>.
\ee

Hence the classical energy-momentum tensor of the system $T_{\mu \nu}$ is defined as $\left<\Psi \right |\hat{T}_{\mu \nu}\left |\Psi \right>=T_{\mu \nu}$.
The semiclassical equation Eq.~(\ref{gr1}) can be obtained from the variational principle \cite{Kibble}
\be\label{quantac}
\delta \left(S_g+S_{\psi}\right)=0,
\ee
where $S_g=\left(1/16\pi G\right)\int{R\sqrt{-g}d^4x}$ is the classical action of the gravitational field, and
\be
S_{\Psi}=\int{\left[{\rm Im}\left \langle \dot{\Psi}|\Psi\right \rangle-\left \langle \Psi |\hat{H}|\Psi \right \rangle +\alpha \left(\left \langle \Psi |\Psi \right \rangle -1\right) \right]dt},
\ee
where $\hat{H}$ is the Hamiltonian operator of the system, and $\alpha $ is a Lagrange multiplier. The variation  of Eq.~(\ref{quantac}) provides the normalization condition for the wave function $\left<\Psi|\Psi\right>=1$, the Sch\"odinger equation for the wave function
\be\label{Sch}
i\left|\dot{\Psi}(t)\right>=\hat{H}(t)\left|\Psi (t)\right>-\alpha (t)\left|\Psi (t)\right>,
\ee
as well as the semiclassical Einstein Eq.~(\ref{gr1}). In this simple case the Bianchi identities require the conservation of the energy-momentum tensor, $\nabla _{\mu}\left<\Psi \right|\hat{T}^{\mu \nu}\left|\Psi \right>=0$.

A very different set of semiclassical Einstein equations can  be obtained by assuming a coupling between the quantum fields and the curvature of the space-time. In the model introduced in \cite{Kibble} the contribution to the total action of the geometry-quantum matter coupling term was assumed to be of the form
\be
\int{RF\left(\left<f(\phi)\right>\right)_{\Psi}\sqrt{-g}d^4x},
\ee
where $F$ and $f$ are arbitrary functions, and $\left(\left<f(\phi)\right>\right)_{\Psi}=\left<\Psi (t)\right|f[\phi (x)]\left|\Psi (t)\right>$. Then, in the presence of such a geometry-matter coupling  the Hamiltonian $H(t)$ in the Schr\"odinger Eq.~(\ref{Sch}) is modified to \cite{Kibble}
\be
\hat{H}(t)\rightarrow \hat{H}_{\Psi}=\hat{H}(t)-\int{N F'\left(\left<f(\phi)\right>\right)_{\Psi}f(\phi)\sqrt{\gamma}d^3\xi},
\ee
where $N$ is the lapse function, $\xi ^i$ are intrinsic coordinates, such that the normal is everywhere time-like, and $\gamma = {\rm det} \;\gamma _{rs}$, where $\gamma _{rs}$ is the metric induced on a surface $\sigma( t )$, which gives a global slicing of the space-time into
space-like surfaces. The effective semiclassical Einstein equation takes the form \cite{Kibble}
\bea\label{123}
R_{\mu \nu}-\frac{1}{2}Rg_{\mu \nu}&=&16\pi G\Big[\left< \hat{T}_{\mu \nu}\right> _{\Psi}+G_{\mu \nu}F
     \nonumber\\
&&-\nabla _{\mu}\nabla _{\nu} F+
g_{\mu \nu}\Box F\Big].
\eea

In Eq.~(\ref{123}) the matter energy-momentum tensor is not conserved, $\nabla _{\mu}\left< \hat{T}^{\mu \nu}\right> _{\Psi}\neq 0$. Thus, this equation describes an effective particle production process, and can be interpreted as giving an effective semiclassical description of the quantum processes in a gravitational field. By modifying the classical part of the gravitational action we can recover the field equations Eqs.~(\ref{field1a}) and (\ref{14}), respectively, used in the present paper. Therefore the physical origin of the matter creation processes considered in the present paper can be traced back to the semiclassical approximation of the quantum field theory in a Riemannian curved geometry.

An interesting and important question is the physical nature of the particles that could be created via gravitationally induced creation processes. The most natural assumption would be that these particles are dark matter particles. It has been conjectured that dark matter may consist of ultra-light particles with masses of the order of $m\approx 10^{-24}$ eV (see \cite{Lee} and references therein). From a physical point of view such a particle may represent a  pseudo Nambu-Goldstone boson. Axions are other ultra-light dark matter candidates, with masses in the range $m\leq 10^{-22}$ eV \cite{Park}. Such extremely very low mass particles can be created even in very weak gravitational fields. An alternative description of dark matter is provided by the so-called scalar field dark matter models \cite{Mat},in which it is assumed  that dark matter is a real scalar field, minimally coupled to gravity, with the mass of the scalar field having a very small value of the order of $m <10^{−21}$ eV. For zero temperature scalar field dark matter models all particles in the system condense to the same quantum ground state, thus forming a Bose-Einstein condensate. Therefore scalar field dark matter models are equivalent to the Bose-Einstein condensate dark matter models \cite{Bose}. This implies, from a physical point of view,  that in the open irreversible thermodynamic  model introduced in the present paper particle creation can take place also in the form of a scalar field. In such a model the evolution of the scalar field dark energy particles, with energy density $\rho _{\phi}$ and pressure $p_{\phi}$, and having a particle number density $n_{\phi}$,  is governed by an equation of the form
\be\label{scalf}
\dot{\rho }_{\phi }+3H\left(\rho _{\phi }+p_{\phi }\right)+\frac{\Gamma _1\left(\rho _{\phi }+p_{\phi }\right)\rho _{\phi }}{n_{\phi }}=0,
\ee
where $\Gamma _1$ is the particle decay rate, determined by the coupling between matter and geometry. For the energy density and pressure of the scalar field dark matter we can assume the standard form
\be
\rho _{\phi}=\frac{\dot{\phi}^2}{2}+U_{int}(\phi ),\qquad p_{\phi }=\frac{\dot{\phi}^2}{2}-U_{int}(\phi ),
\ee
where $U(\phi)$ is the scalar field self-interaction potential.

The creation pressure corresponding to the scalar field creation processes can be obtained as
\be
p_c^{(\phi )}=\frac{\Gamma _1\left(\rho _{\phi }+p_{\phi }\right)\rho _{\phi }}{3Hn_{\phi }}.
\ee

 It is interesting to note that Eq.~(\ref{scalf}), which describes the creation of a scalar field as a result of the geometry-matter coupling, can be written in an equivalent form as
\be\label{scaleq}
\ddot{\phi }+3H\dot{\phi }+\Gamma \left(\phi, \dot{\phi },U\right)\dot{\phi }+U_{int}'(\phi )=0,
\ee
where we have denoted $\Gamma \left(\phi, \dot{\phi }, U\right)=\Gamma _1\rho _{\phi }/n_{\phi }$. Therefore in the scalar field dark matter model a friction term in the scalar field evolution equation Eq.~(\ref{scaleq}) does appear  naturally, and in a general form, as a direct consequence of the  irreversible thermodynamics of open systems as applied to the dark matter case. Hence scalar field dark matter can be a result of the cosmological particle production due to the geometry-matter coupling in modified gravity theories. For gravitational models with an action given by an arbitrary function of the Ricci scalar, the matter Lagrangian density, a scalar field and a kinetic term constructed from the gradients of the scalar field, respectively, see \cite{Min}.

The $Lf\left(R,L_m\right)$ gravitational theory investigated in the present
paper predicts the possibility that matter creation, associated with the
curvature-matter coupling, could also occur in the present-day universe, as
proposed by Dirac \cite{22} a long time ago. The late expansion of the
Universe \cite{1,2} may be considered as an empirical evidence for matter
creation, and a viable alternative to the mysterious dark energy. Presently
the existence of some forms of the curvature-matter coupling leading to
matter creation processes cannot be fully ruled out by the existing
cosmological observations or by astrophysical data. Presumably, the
functional forms of the potentials $V(\psi)$ and $U(\psi)$ that completely
characterize the $Lf\left(R,L_m\right)$ gravitational theory will be
provided by fundamental quantum field theoretical models of the
gravitational interaction, thus opening the possibility of an in depth
comparison of the predictions of the $Lf\left(R,L_m\right)$ gravity with
cosmological and astrophysical observational data.

\section*{Acknowledgements}
FSNL is supported by a Funda\c{c}\~{a}o para a Ci\^{e}ncia e Tecnologia
Investigador FCT Research contract, with reference IF/00859/2012, funded by
FCT/MCTES (Portugal). FSNL and JPM acknowledge financial support of the Funda%
\c{c}\~{a}o para a Ci\^{e}ncia e Tecnologia through the grant
EXPL/FIS-AST/1608/2013 and UID/FIS/04434/2013.
 DP was partially supported by the ``Ministerio de Econom\'{\i}a y Competitividad, Direcci\'{o}n General de Investigaci\'{o}n  Cient\'{\i}fica y T\'{e}cnica",  Grant N$_{0}$. FIS2012-32099.



\end{document}